\documentclass [11pt]{article}
\usepackage {graphicx}
\usepackage {color}
\begin{document}
\begin{center}
\textbf{ABOUT HIERARCHY OF MULTIFREQUENCY QUASIPERIODICITY REGIMES IN DISCRETE LOW-DIMENSIONAL KURAMOTO MODEL}
\end{center}

\begin{center}
ALEXANDER P. KUZNETSOV and YULIYA V. SEDOVA

Kotel'nikov's Institute of Radio-Engineering and Electronics of RAS, Saratov
Branch
Zelenaya 38, Saratov, 410019, Russian Federation

apkuz@rambler.ru, sedovayv@yandex.ru
\end{center}

\bigskip

A dynamics of a low-dimensional ensemble consisting of connected in a
network five discrete phase oscillators is considered. A two-parameter
synchronization picture which appears instead of the Arnold tongues with an
increase of the system dimension is discussed. An appearance of the Arnol'd
resonance web is detected on the ``frequency -- coupling'' parameter plane.
The cases of attractive and repulsive interactions are discussed.


\section{Introduction}

Investigation of ensembles of interacting oscillators is an important
problem with applications in various fields such as radiophysics, laser
physics, biophysics, dynamics of gene networks, etc. \cite{Pikovsky,Landa,Balanov,Glass,Kuramoto}. Traditionally,
the Kuramoto model which is used for this purposes represents a set of
globally coupled phase oscillators \cite{Pikovsky,Landa,Kuramoto,Strogatz,Acebron,Maistrenko}. The main effect observed in
this system is an appearance of a coherent state in the medium field
generated by the ensemble (Kuramoto transition). However, the medium field
model is efficient in case when the network contains a very large number of
oscillators. At the same time, it is interesting to study a behavior of
low-dimensional ensembles containing a relatively small number of
oscillators. It is important for various applications, for example in
biophysics when several subsystems with different natural rhythms interact
with each other. This problem is also fundamental in the following context.
Each new element addition into the system adds a new frequency. As a result,
an emergence of high-dimensional quasiperiodic oscillations associated with
the multi-dimensional invariant tori becomes possible. Variation of at least
one fundamental frequency may initiate various resonant conditions. Due to
the coupling of type ``each-to-each'', the number of such resonances is
maximal in the network of oscillators. With an increase in the coupling,
hierarchy of resonances is observed. This is a situation when
low-dimensional tori arise on the surfaces of high-dimensional tori. One can
expect a complicated structure of such resonances. Thus, this problem may be
interpreted as a generalization of the Arnol'd tongues to the
multi-frequency systems.

At transition to systems with multi-frequency quasiperiodicity, however, we
have to apply new methods of analysis. Indeed, it is necessary to identify
quasiperiodic regimes of different dimensions and details of the internal
organization of the domains of their existence. This problem can be solved
by using a Lyapunov analysis, which should be performed at each point of the
parameter plane. Then the analysis of the spectrum of Lyapunov exponents
enables us to determine the type of regime. However, a solution of such a
problem leads to the following difficulty. With an increase in number of
oscillators a computation time required for the regime definition at each
point in the parameter space becomes very large. This can be partially
overcome by using a simple model and transition from the flow systems to the
maps. The simplest method of map constructing is to replace time derivatives
by finite differences in the dynamic equations. This approach is used, for
example, in the conservative chaos theory (Chirikov--Taylor map) \cite{Zaslavsky}, in the
constructing of ``predator--sacrifice'' discrete models \cite{Ghaziani,Han}, the
simplest variants of genetic networks (Andrecut--Kauffman map) \cite{deSouza,Andrecut}, in
the analysis of normal forms of some bifurcations (Bogdanov map) \cite{Arrowsmith}, etc.

It should be noted, however , that the relationship between continuous and
discrete model is quite a tricky question\footnote{See, in this connection,
e.g., \cite{deLima}.}. The dynamics of the discrete model inherits partly the
properties of the prototype system, but in many ways is more rich in regard
of possible nonlinear phenomena. This is true even for the simplest case
of two-element systems when discretization procedure leads to a transition
from the classical Adler equation to the sine circle map \cite{Pikovsky,Landa,Balanov}. Increasing
the number of oscillators leads to the need to move from circle map to a map
defined on the torus of sufficiently high dimension.

This approach is sufficiently constructive for studying networks consisting
of elements with complex dynamics. For example, it was used for networks
with large number of elements in recent paper \cite{Barlev}. Low-dimensional discrete
networks with three and four elements have been investigated in \cite{Vasylenko} and
\cite{Maistrenko10}, respectively. The case of three oscillators is relatively simple. One
of the main results is an appearance of the threshold coupling value for the
domain of the complete synchronization between all oscillators \cite{Vasylenko}. For the
four interacting oscillators a problem is much more complicated. In \cite{Maistrenko10},
the authors have focused on the situation of spectrum with equidistant
frequency pattern. They have introduced a single frequency parameter which
determines a frequency detuning for all oscillators. Thus, a possibility of
various resonances in the system is significantly weakened in this
formulation of the problem.

In this paper, we investigate an ensemble of five globally coupled discrete
phase oscillators. We consider the spectrum with non-equidistant frequency
pattern. Four frequencies remain constant and one frequency is varied. In
this approach, oscillator with varying frequency may be in resonance with
any of the other ones or with some clusters in the network.

With this approach, the selected oscillator can turns out to be at a
resonance with any of the remaining oscillators, or with some clusters in
the network. Thus a scan is performed of network properties via sweeping the
frequency of the "test" oscillator. This method of "testing oscillator" may
be promising and subsequently extrapolated to more complex network of a
large number of oscillators.

\section{Two-parameter investigation of oscillation regimes}
Let us consider a network with five discrete phase oscillators
\begin{equation}
\label{eq1}
\psi _{n} \to \omega _{n} + \psi _{n} + \mu {\sum\limits_{i = 1}^{5} {\sin
(\psi _{i} - \psi _{n} )}} .
\end{equation}
Here $\psi _{i} $ is the phase of $i$-th oscillator, $\omega _{i} $ is its
natural frequency, $\mu $ is the coupling parameter.

A dimension of the system (\ref{eq1}) can be lowered by one. For this purpose we
introduce relative phases of the oscillators
\begin{equation}
\label{eq2}
\theta = \psi _{2} - \psi _{1} {\rm ,}
\quad
\varphi = \psi _{3} - \psi _{2} {\rm ,}
\quad
\alpha = \psi _{4} - \psi _{3} {\rm ,}
\quad
\beta = \psi _{5} - \psi _{4} {\rm .}
\end{equation}

Let us define difference frequencies $\Delta _{i} $ relative to the first
oscillator, i.e. $\Delta _{i - 1} = \omega _{i} - \omega {}_{1}$. Thus, we
obtain:
\begin{equation}
\label{eq3}
\begin{array}{l}
 \theta \to \Delta _{1} + \theta + \mu [ - 2\sin \theta + \sin \varphi +
\sin (\varphi + \alpha ) + \sin (\varphi + \alpha + \beta ) - \sin (\theta +
\varphi ) - \\
 \,\,\,\,\,\,\,\,\,\, - \sin (\theta + \varphi + \alpha ) - \sin (\theta +
\varphi + \alpha + \beta )], \\
 \varphi \to \Delta _{2} - \Delta _{1} + \varphi + \mu [ - 2\sin \varphi +
\sin \theta + \sin \alpha - \sin (\theta + \varphi ) - \sin (\varphi +
\alpha ) + \\
 \,\,\,\,\,\,\,\,\,\, + \sin (\alpha + \beta ) - \sin (\varphi + \alpha +
\beta )], \\
 \alpha \to \Delta _{3} - \Delta _{2} + \alpha + \mu [ - 2\sin \alpha + \sin
\varphi + \sin \beta + \sin (\theta + \varphi ) - \sin (\varphi + \alpha ) -
\\
 \,\,\,\,\,\,\,\,\,\, - \sin (\alpha + \beta ) - \sin (\theta + \varphi +
\alpha )], \\
 \beta \to \Delta {}_{4} - \Delta _{3} + \beta + \mu [ - 2\sin \beta + \sin
\alpha + \sin (\varphi + \alpha ) - \sin (\varphi + \alpha + \beta )- \\
 \,\,\,\,\,\,\,\,\,\, - \sin
(\alpha + \beta )+ \sin (\theta + \varphi + \alpha ) - \sin (\theta +
\varphi + \alpha + \beta )]. \\
 \end{array}
\end{equation}

Choose the natural frequencies of the described system in the following way.
Eqs.~(\ref{eq3}) contain only differences between natural frequencies. Therefore,
frequency of one of the oscillators (e.g., of the first one) can be fixed.
We fix also frequencies of the third, fourth and fifth oscillators in such a
way that the frequency parameters have unequal values: $\Delta _{2} =
0.1,\Delta _{3} = 0.45$, $\Delta _{4} = 1$. Frequency of the second
oscillator will be varied by changing of the frequency parameter $\Delta _{1}
$.

For analyzing of system (\ref{eq3}) we use the method of the charts of Lyapunov
exponents \cite{Baesens,Broer,Emelianova,Kuznetsov,Khibnik}. According to this method, first we select a point in the
parameter plane and calculate all Lyapunov exponents of system (\ref{eq3}). Then we
color this point in accordance with the type of a regime. In such a way we
scan the entire parameters plane in the selected range. Fig.~1 shows the
chart of Lyapunov exponents in the plane of the coupling parameter $\mu $
versus the second oscillator frequency $\Delta _{1} $. The general view of
the chart is presented in Fig.~1a, and Fig.~2b gives its fragment which
illustrates the basic multi-frequency resonances. The color palette is
specified in the figure caption, so that we can reveal:

$\bullet$ periodic attractor $P$ with all the negative Lyapunov exponents;

$\bullet$ two-frequency regime $T_{2} $ with zero Lyapunov exponent;

$\bullet$ three-frequency regime $T_{3} $ with two zero Lyapunov exponents;

$\bullet$ four-frequency regime $T_{4} $ with three zero Lyapunov exponents;

$\bullet$ five-frequency regime $T_{5} $ with four zero Lyapunov exponents;

$\bullet$ chaotic regime $C$ with positive largest Lyapunov exponent;

$\bullet$ hyperchaos regime $H$ with at least two positive Lyapunov exponents.

For simplicity, let us call multi-frequency quasiperiodic regimes as domains
of tori with corresponding dimension. But formally, for the reduced phase
system~(\ref{eq3}) an invariant hypersurfaces are realized.
\begin{figure}[h!]
\begin{center}
\includegraphics[height=8cm, keepaspectratio]{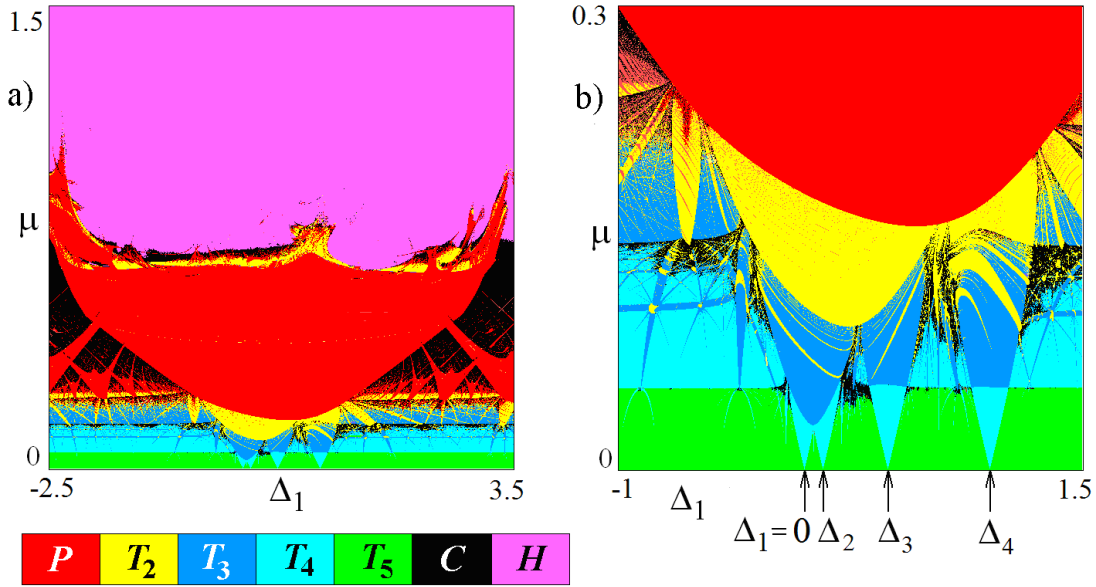}
\end{center}
\caption{Chart of Lyapunov exponents for the network of five phase
oscillators (\ref{eq3}) and its enlarged fragment for $\Delta _{2} = 0.1,\Delta
_{3} = 0.45$, $\Delta _{4} = 1$.}
\label{fig1}
\end{figure}
\begin{figure}[h!]
\begin{center}
\includegraphics[height=9cm, keepaspectratio]{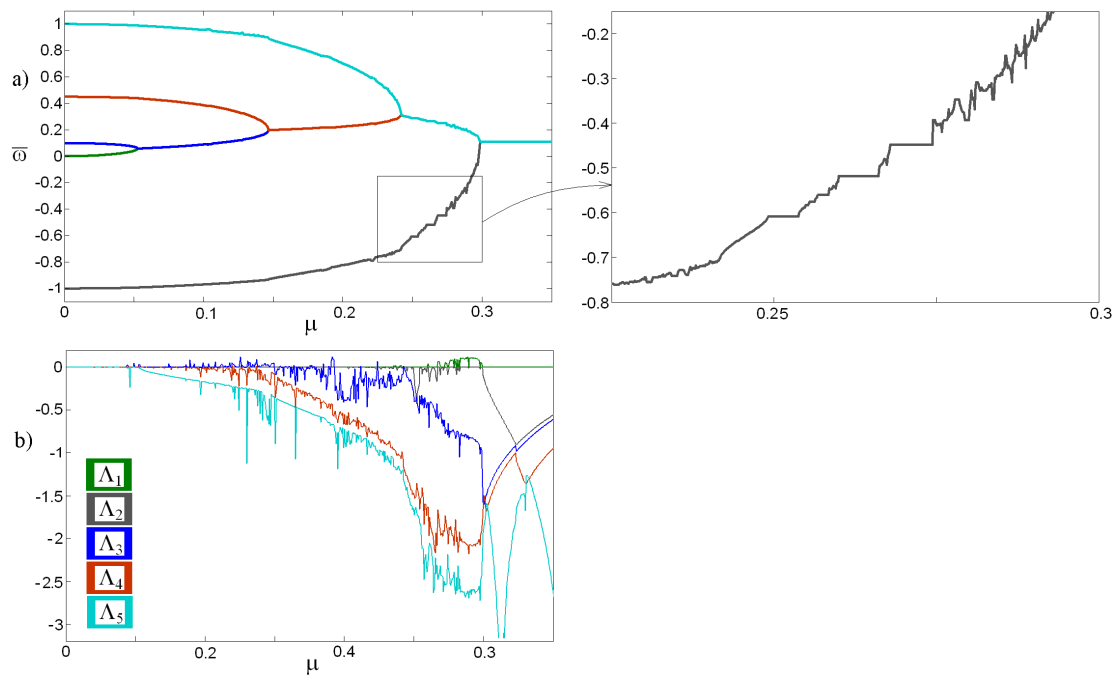}
\end{center}
\caption{(a) Synchronization tree and its enlarged fragment (right)
illustrating complicated structure of this tree. (b) Plot of Lyaponov
exponents for the network of five discrete phase oscillators (\ref{eq3}). Values of
the parameters are\textbf{} $\Delta _{1} = - 1$, $\Delta _{2} = 0.1,_{{\rm
}}\Delta _{3} = 0.45$.}
\label{fig2}
\end{figure}

Fig.~2a shows the dependence of frequencies $\bar {\omega} _{i} = {\mathop
{\lim} \limits_{n \to \infty} } {\frac{{\psi _{i}^{(n)} - \psi _{i}^{(0)}
}}{{n}}}$ which are observed in the system on coupling value. Consider the
case $\Delta _{1} = - 1$ (left border in Fig.~1b). When $\mu = 0$, the
observed frequencies are equal to the natural frequencies of the
oscillators, i.e., according to (\ref{eq3}):
\begin{equation}
\label{eq4}
\bar {\omega} _{1} = 0{\rm ,}
\quad
\bar {\omega} _{2} = \Delta _{1} {\rm ,}
\quad
\bar {\omega} _{3} = \Delta _{2} {\rm ,}
\quad
\bar {\omega} _{4} = \Delta _{3} {\rm ,}
\quad
\bar {\omega} _{5} = \Delta _{4} .
\end{equation}

In Fig.~2a one can see a sequential emergence of clusters which corresponds
to the consecutive merging of the ``tree branches''. For example, the first
cluster arises when the first and the second oscillators join together.
After this, the third one is joined to them, and etc. It should be noted
that the number of clusters does not strictly correspond to the dimension of
the observed torus. A plot of Lyapunov exponents in Fig.~2b verifies this
fact. An enlarged fragment of the tree in Fig.~2a (right) shows that the
tree branch is jagged in the domain where there is multiple alternation of
two-frequency, periodic and chaotic regimes.

Now discuss structure of the parameter plane given in Fig.~1. Consider the
domain of small coupling values. One can see that the five-frequency tori
are dominant. However, there are several tongues of the four-frequency
regimes. At the bottom, they look like traditional Arnol'd tongues. At some
points they touch the $\Delta _{1} $-axis. We can derive coordinates of
these points using physical consideration. Variation of the frequency
$\omega {}_{1}$ leads to the consistent resonances between the second
oscillator and the first, third, fourth and fifth one. From the definition
of frequency detunings $\Delta _{i-1} = \omega _{i} - \omega {}_{1}$, we
obtain four resonance conditions
\begin{equation}
\label{eq5}
\Delta _{1} = 0{\rm ,}
\quad
\Delta _{1} = \Delta _{2} {\rm ,}
\quad
\Delta _{1} = \Delta _{3} {\rm ,}
\quad
\Delta _{1} = \Delta _{4} {\rm .}
\end{equation}
These values are indicated by vertical arrows in Fig.~1b.

We choose the spectrum of oscillators so that the frequencies of the fourth
oscillator and the fifth one are sufficiently different. Thus, the two
right-hand tongues in Fig.~1b satisfying the conditions $\omega _{4} \approx
\omega {}_{1}$ and $\omega _{5} \approx \omega {}_{1}$ have a very simple
structure. They represent two domains of four-frequency tori inside the
five-frequency one. In these cases, the second oscillator is in resonance
with only one oscillator, and the other ones are actually independent. This
is a kind of the ``individual'' resonance.

At the same time, frequencies of the first and the third oscillators are
nearly equal: $\omega _{1} \approx \omega {}_{3}$ ($\Delta _{2} \approx 0$).
Therefore, the second test oscillator interacts with this pair at the
variation of its frequency $\Delta_{1}$. This is a kind of the
``collective'' resonance. In such a case synchronization picture is more
complicated and is shown in detail in Fig.~3a. One can see that the
four-frequency tongues are closed by edges forming a characteristic oval
domain of three-frequency tori $T_{3} $. In this case, a cluster of three
oscillators (the first, second and third ones) may arise.

At the same time, inside the five-frequency domain $T_{5} $ there are many
thin higher-order tongues of four-frequency tori. Fig.~3b illustrates two
types of them. In the first case, the tongue has a traditional cusp shape.
In the second case, a fan-shaped system of four-frequency tori is located at
the bottom of the rounded three-frequency domain. Small chaotic regions may
also be observed.
\begin{figure}[h!]
\begin{center}
\includegraphics[height=7.5cm, keepaspectratio]{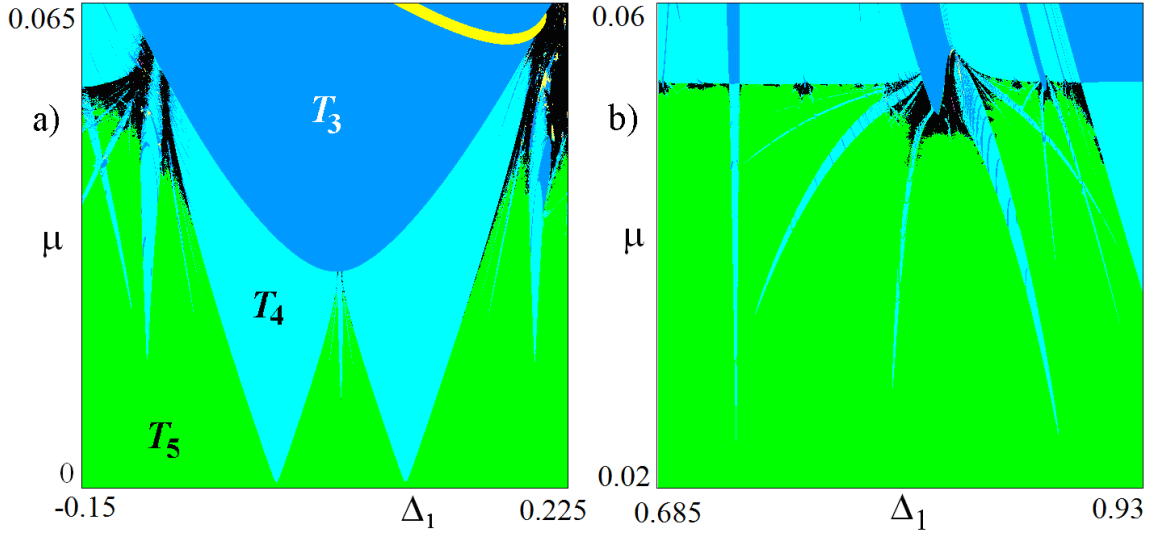}
\end{center}
\caption{(a) Integration of the four-frequency tongues.
(b) Two higher-order tongues of four-frequency
tori. The colour palette is similar to that in Fig.~1.}
\label{fig3}
\end{figure}

With an increase in the coupling parameter ($\mu \ge 0.05$), the
five-frequency regimes in Fig.~1 are replaced by the four-frequency regimes.
The corresponding boundary looks like a horizontal line. It is a saddle-node
bifurcation line for the four-frequency torus. Above this line, there is a
characteristic picture of the Arnol'd resonance web \cite{Broer} which is shown in Fig.~4a. In this case,
there is a network of three-frequency domains with dual-frequency regimes
arising at their regions of intersection. It should be noted that this
result is somewhat unexpected. The resonance web arises usually on the
parameter plane of the fundamental frequencies of the oscillators \cite{Baesens,Broer}.
In our case, one of the parameters is the coupling. We can give following
explanation of this fact. The oscillation frequency for the arising cluster
depends on the coupling value. Thus, the resonance conditions occur with
simultaneous variation of the frequency and the coupling. Note that with an
increase in coupling $\mu$ the resonance web structure remains in Fig.~4a,
but the four-frequency regimes are replaced by the chaotic ones.
\begin{figure}[h!]
\begin{center}
\includegraphics[height=7cm, keepaspectratio]{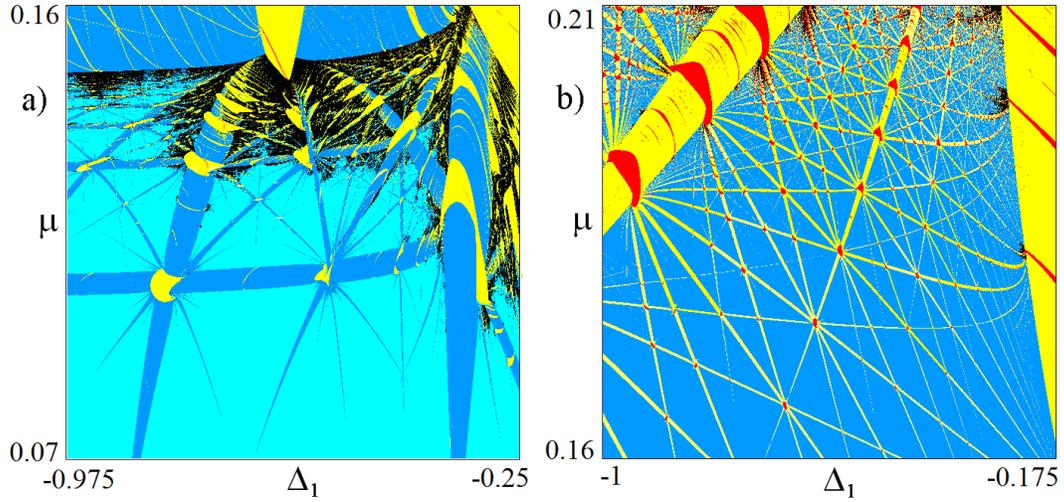}
\end{center}
\caption{Resonance web for the network of five discrete phase
oscillators (\ref{eq3}) inside the domain of (a) four-frequency and (b)
five-frequency tori.}
\label{fig4}
\end{figure}

For $\mu \ge 0.15$, the three-frequency region is observed. In this case,
there is also a resonance web, but it is on the basis of the three-frequency
regimes as is shown in Fig.~4b.

For larger values of $\mu $, there is a domain of two-frequency
quasiperiodic regimes which looks like a band (Fig.~5a). There is also a
system of higher-order complete resonances. They look like the traditional
Arnol'd tongues, but with the destroyed bottoms. One can see diverse kinds
of the complete resonances on the chart of periodic regimes in Fig.~5b. In
this case, the different colours indicate different periods of cycles of the
analyzed map. Near the bottom of each tongue of periodic regimes, the system
of fan-shaped domains of two-frequency tori arises. Inside the tongues of
periodic regimes, there is a period doubling which leads to chaos with an
increase in the coupling parameter.
\begin{figure}[h!]
\begin{center}
\includegraphics[height=8.5cm, keepaspectratio]{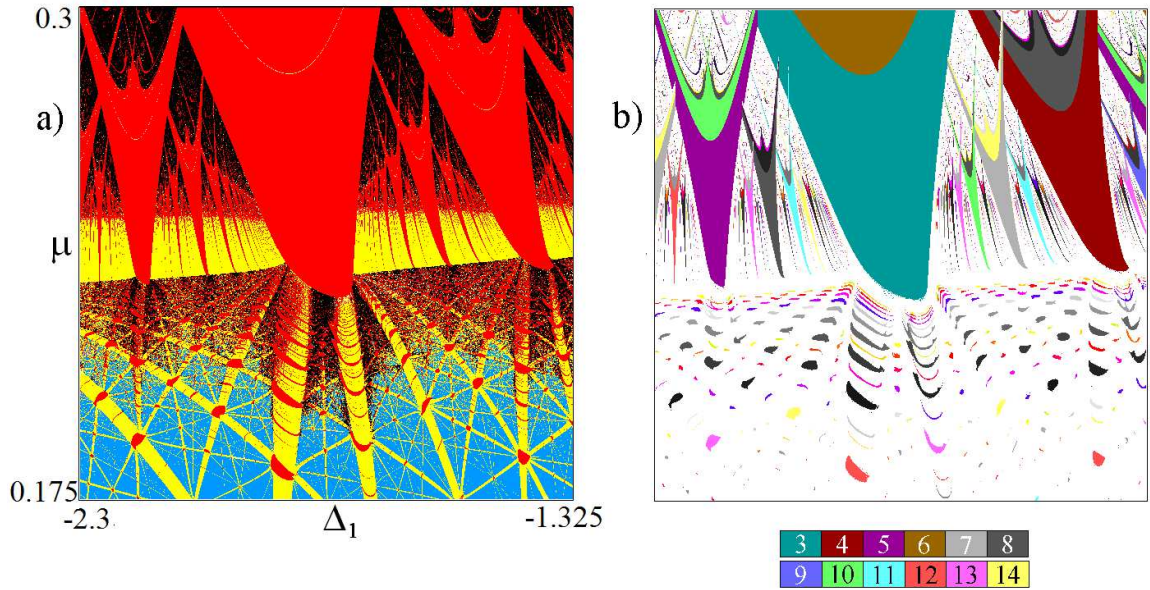}
\end{center}
\caption{Higher-order tongues of the complete synchronization
embedded in the quasiperiodic domains of different dimension and chaos. (a)
Chart of Lyapunov exponents; (b) chart of periodic regimes; numbers below
indicate periods of cycles.}
\label{fig5}
\end{figure}

\section{Comparison with the dynamics of the chain of oscillators}
Now discuss an influence of the coupling geometry in the system to the
synchronization. For this purpose, we compare the above results with the
results for oscillators coupled in a linear chain. In this case, the phase
equations are
\begin{equation}
\label{eq6}
\begin{array}{l}
 \psi _{1} \to \omega {}_{1} + \psi _{1} + \mu \sin (\psi _{2} - \psi _{1}
), \\
 \psi _{2} \to \omega _{2} + \psi _{2} + \mu {\left[ {\sin (\psi _{1} - \psi
_{2} ) + \sin (\psi _{3} - \psi _{2} )} \right]}, \\
 \psi _{3} \to \omega _{3} + \psi _{3} + \mu {\left[ {\sin (\psi {}_{2} -
\psi _{3} ) + \sin (\psi _{4} - \psi _{3} )} \right]}, \\
 \psi _{4} \to \omega _{4} + \psi {}_{4} + \mu {\left[ {\sin (\psi _{3} -
\psi _{4} ) + \sin (\psi _{5} - \psi _{4} )} \right]}, \\
 \psi _{5} \to \omega _{5} + \psi _{5} + \mu \sin (\psi _{4} - \psi _{5} ).
\\
 \end{array}
\end{equation}

For the relative phases (\ref{eq2}), we obtain the following system of equations:
\begin{equation}
\label{eq7}
\begin{array}{l}
 \theta \to \theta + \Delta _{1} + \mu ( - 2\sin \theta + \sin \varphi ), \\
 \varphi \to \varphi + \Delta _{2} - \Delta _{1} + \mu ( - 2\sin \varphi +
\sin \theta + \sin \alpha ), \\
 \alpha \to \alpha + \Delta _{3} - \Delta _{2} + \mu ( - 2\sin \alpha + \sin
\beta + \sin \varphi ), \\
 \beta \to \beta + \Delta _{4} - \Delta _{3} + \mu ( - 2\sin \beta + \sin
\alpha ). \\
 \end{array}
\end{equation}

Fig.~6 shows the corresponding chart of Lyapunov exponents constructed for
the same values of the parameters as in Fig.~1. In this case, the number of
basic resonance tongues of four-frequency tori reduces from four to two.
This has a physical explanation. Indeed, variation of the frequency for the
second oscillator $\omega _{2} $ may lead to one of the two possible
resonances $\omega _{2} = \omega _{1} $, $\omega _{2} = \omega _{3} $due to
the coupling geometry in the chain. This provides the conditions
\begin{equation}
\label{eq8}
\Delta _{1} = 0{\rm ,}
\quad
\Delta _{1} = \Delta _{2} {\rm .}
\end{equation}
Thus, we obtain an interesting result: the number of basic tongues of
four-frequency tori equals to the number of nearest neighbors of the
oscillator with a variable frequency. It will also be true for networks with
a more complex coupling topology.

\begin{figure}[h!]
\begin{center}
\includegraphics[height=7.5cm, keepaspectratio]{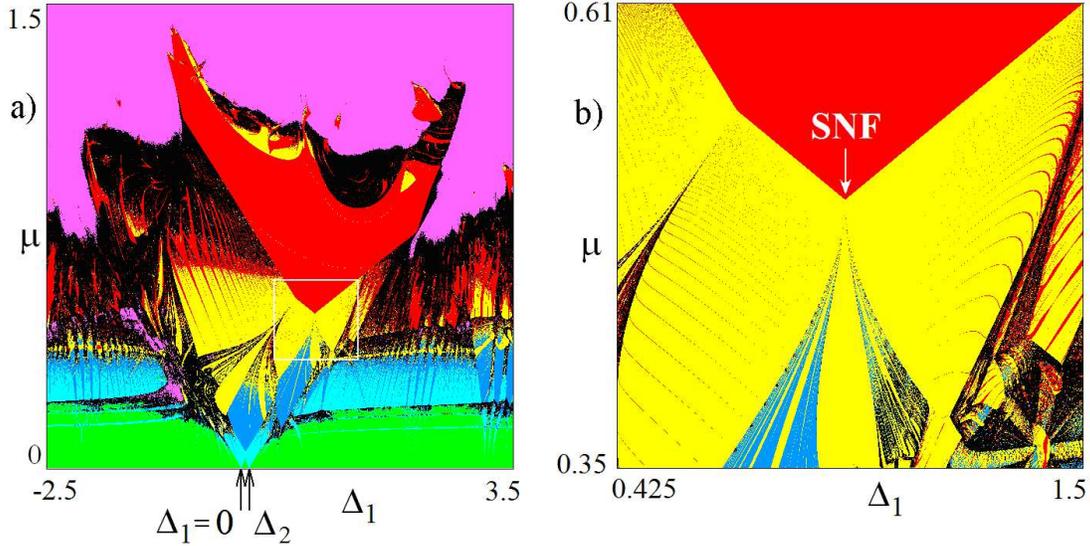}
\end{center}
\caption{(a) Chart of Lyapunov exponents for the chain of five phase
oscillators (\ref{eq7}) and (b) its enlarged fragment. Values of the parameters are
$\Delta _{2} = 0.1,\Delta _{3} = 0.45$, $\Delta _{4} = 1$. SNF is saddle-node fan  point  \cite{Baesens,Emelianova}.}
\label{fig6}
\end{figure}

Another conclusion is that the shape of the complete synchronization domain
for the network and for the chain of oscillators is different (compare
Fig.~1b and Fig.~6b). In the last case, there are characteristic angles
corresponding to the codimension-two points (Fig.~6b). These points are
typical also for flow models \cite{Emelianova}.

\section{System with anti-phase synchronization}
We have considered the dissipative coupling which tends to equalize the
states of oscillators. In the simplest case of two elements, this type of
coupling leads to the in-phase synchronization. However, the case of
negative values of the coupling is also important. In this
case, the anti-phase synchronization is stable for
two coupled elements. This type of coupling is characteristic, for example,
for laser physics when lasers are optically coupled by radiation through the
sidewalls of waveguides \cite{Khibnik,Glova02,Glova03}. It may be called an active coupling or
repulsive interaction \cite{Hong}.

Let us discuss the case of an active coupling. Firstly, consider a chain of
oscillators. In this case, there is a certain symmetry in the system.
Indeed, Eqs.~(\ref{eq7}) are invariant under a linear change of variables
\begin{equation}
\label{eq9}
\mu \to - \mu {\rm ,}
\quad
\theta \to \theta + \pi ,
\quad
\varphi \to \varphi + \pi ,
\quad
\alpha \to \alpha + \pi ,
\quad
\beta \to \beta + \pi .
\end{equation}
This transformation of variables does not change the type and stability
characteristics of fixed points. Only a phase shift of $\pi $ occurs and a
change from the in-phase to anti-phase regimes is observed. Therefore, the
chart of dynamic regimes for the oscillators with anti-phase synchronization
looks exactly like the chart in Fig.~6. Thus, a complete synchronization
between all the subsystems is possible in the chain of coupled oscillators.
\begin{figure}[h!]
\begin{center}
\includegraphics[height=8cm, keepaspectratio]{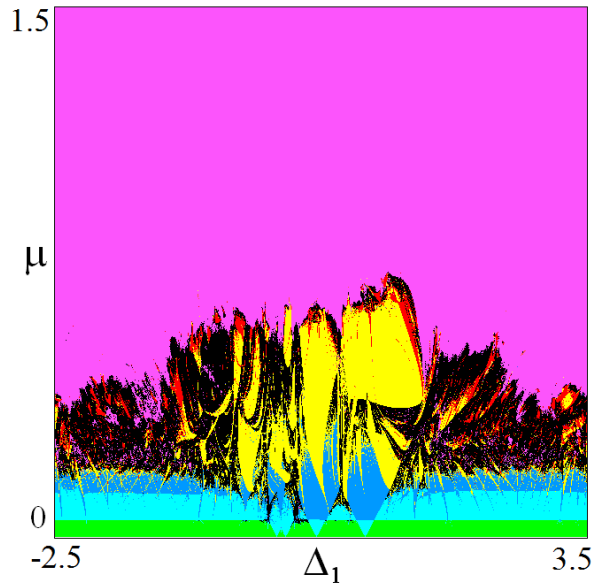}
\end{center}
\caption{Chart
of Lyapunov exponents for the network of five phase oscillators in case of
an active coupling for $\Delta _{2} = 0.1, \Delta _{3} = 0.45, \Delta
_{4} = 1$.}
\label{fig7}
\end{figure}

There is another situation in case of globally coupled elements. In this
case, Eqs.~(\ref{eq3}) are not invariant under the change of variables (\ref{eq8}) due to
the presence of terms containing the sum of relative phases. Therefore, a
structure of the parameter space for the network of such elements differs
from that for the chain. Fig.~7 shows the chart of Lyapunov exponents for
the oscillators with anti-phase synchronization with the same values of
fundamental frequencies as in Fig.1. One can see that for small values of
$\mu $, the picture is partly equivalent to the case of the dissipative
coupling. Thus, if two or three oscillators are captured inside the network
of coupled oscillators, the system behaviour is qualitatively the same for
any sign of the coupling parameter. For high values of the coupling, there
are great differences between these two situations. An ordered structure of
two-frequency quasiperiodic domains, which is typical for the dissipative
coupling, is destroyed. A domain of the complete synchronization virtually
disappears in case of an active coupling and only a few isolated ``islands''
of periodic regimes are visible. Thus, the Kuramoto transition does not
occur for an active coupling. Instead, the chaotic or even hyperchaotic
regimes are observed\footnote{Such result is in accordance with the
numerical simulation for the network of a large number of oscillators \cite{Hong}.}. This result is important for laser arrays because it means that the
coupling configuration of type ``each-to-each'' is not always a good method
to get coherent radiation.

\section{Conclusion}
Two-parameter Lyapunov analysis is an effective tool for the study of
ensembles of discrete phase oscillators. Another useful trick is a frequency
scanning of the system properties via using one dedicated oscillator.
Features of the dynamics in low-dimensional network ensembles are
investigated in case of the five discrete phase oscillators. There are
characteristic domains of different-order resonance tori. Bottoms of these
domains are destroyed and form the fan-shaped system of domains of
higher-order tori. For medium values of the coupling, there is Arnol'd
resonance web on the ``frequency detuning -- coupling'' parameter plane. It
exists in both the domains of three- and four-frequency tori. The case of
synchronization of oscillators with repulsive interaction in the domain of
strong coupling is significantly different from the case of in-phase
synchronization, in particular, full synchronization modes are atypical.

\section*{Acknowledgments}
The work was supported by grant of the President of the Russian Federation for state support of Leading Scientific Schools NSH-1726.2014.2. Yu.V.S. acknowledges support from Russian Foundation for Basic Research (grant No 14-02-31064).

\end{document}